\documentclass{article}

\usepackage{slashed}
\usepackage{multicol,multirow}

\usepackage{pgfplots}
\usepackage{tikz}
\usetikzlibrary{patterns}

\usepackage{float}

\usepackage{paralist}

\usepackage{graphicx}
\usepackage{caption}
\usepackage{subcaption}

\usepackage{hyperref}
\usepackage{amsmath,amsfonts,amssymb}
\usepackage[all]{xy}
\usepackage{breqn}

\usepackage{xcolor}
\usepackage{array}
\title{Hybrid stars within the framework of the Sigma-Omega-Rho model combined with the MIT and NJL models}
\author{Reza Karimi\thanks{Corresponding author:\href{mailto:rezakarimi0225@gmail.com}{rezakarimi0225@gmail.com}}, H.R. Moshfegh\thanks{\href{mailto:hmoshfegh@ut.ac.ir}{hmoshfegh@ut.ac.ir}}\\Department of Physics, University of Tehran,\\ Post Office Box 14395-547, Tehran, Iran}

\date{}

\begin{document}
    \maketitle
    \begin{abstract}
    	In this paper, we investigate the structure of hybrid stars consisting of hadrons (neutrons, protons, sigmas, lambdas), leptons (electrons, muons), and quarks (up, down, strange). We use a relativistic mean-field (RMF) model namely the Sigma-omega-rho model for the hadronic phase and the MIT bag model as well as the NJL model for the quark phase. In addition, Maxwell and Gibbs conditions are employed to investigate the hadron-Quark phase transition. Finally, by obtaining the mass-radius relation, $ M (M_{sun}) \leqslant 2.07 $ is predicted for such hybrid stars.
    \end{abstract}
    

	\section{Introduction}
	Compact objects (neutron stars) are of crucial importance in studying matter under extreme conditions \cite{e001}. It seems that our knowledge of neutron star matter (especially, the core of neutron star matter) is not sufficient to predict some observational constraints on the mass and radius of neutron stars. Up to now, there is no unique equation of state (EOS) for the description of neutron star matter in all possible densities. Also, the composition of neutron star matter is in doubt, whether hyperons and quarks can be present in the core of neutron stars.
	
	Recently, the most massive neutron star (NS) known as \textbf{PSR J0952-0607} was discovered with a mass \textbf{2.35} times as much as the Sun \cite{r01}. Also, earlier neutron stars like \textbf{PSR J0740+6620} with masses higher than two (times as much as the Sun) have been discovered \cite{r02}. Therefore, theoretical modeling of neutron star that can predict a mass above two ($ M_{sun} $) for the star by considering the real structure of neutron stars is important.
	
	As the density and/or temperature increase in the center of neutron stars, hadronic nuclear matter undergoes a deconfinement transition to quark phase, consisting of quarks and gluons rather than of separate hadrons \cite{r03,r04}. In this case, the neutron star is known as the hybrid star. Investigating the probability of phase transition from hadronic phase to quark phase in the center of the hybrid star and predicting the maximum mass of the hybrid star is a challenging issue in the field.
	
	There are different models to describe the equation of states of the hadronic phase. For example, the Fermi hypernetted chain (FHNC) theory \cite{r04fe}, the lowest-order constraint variational (LOCV) method \cite{er04,r6}, Methods based on Green's function, the Brueckner-Hartree-Fock (BHF) many-body theory \cite{04000fe} or mean field models such as the quark-meson coupling (QMC) model \cite{r05,e02}, and relativistic mean-field (RMF) models are used to describe the hadronic phase. There are different types of RMF models such as TM1 \cite{z1}, and IUFSU \cite{z2}. In this case, we employ an RMF model namely the Sigma-omega-rho model - with correction on the hyperon coupling constants - to describe the hadronic phase \cite{r3}. In this model, calculations of including the distinct hyperons in the equation of state (EOS) are much easier than in other types.
	
	Also, there are different models to describe quark matter such as the mass density dependent (MDD) model \cite{r05fe} and the field correlator method (FCM) \cite{r05fe1}. However, the MIT bag \cite{r06,e03} and Nambu-Jona-Lasinio (NJL) \cite{r07,e04} models are employed to clarify quark matter in this study. In the case of hybrid stars, the phase transition is discussed with the help of Maxwell and Gibbs criteria. The Maxwell construction creates a sharp hadron-to-quark phase transition and has been broadly utilized in later research \cite{fe01,fe02,fe03,fe04}. However, the Gibbs construction describes a blended soft phase of hadrons and quarks \cite{fe11}, and has been studied to explore the recent detections of gravitational-wave \cite{fe12}. Either way, comparing their predictions can provide valuable physical knowledge. 
	
This article is arranged as follows. In Section \ref{sec2}, we briefly introduce the models used for the hadronic and the quark phases. In section \ref{sec3}, first, the method of phase transition from the hadronic phase to the quark phase is discussed. Then the results related to the equation of states of the hybrid star under different phase transitions are given. In the last part of section \ref{sec3}, the results related to the mass-radius relation of the hybrid star are reported. Finally summary and discussion is presented in section \ref{sec4}.
	\section{Formalism}\label{sec2}
		\subsection{Hadronic phase}
		The hadronic phase is characterized by the Sigma-omega-rho model. In this model, the interactions  describe as the exchange of  $\sigma$, $\omega$ and $\rho$ mesons in the mean-field approximation. Based on this model, the Lagrangian could be written as below \cite{r1,r2},
		\begin{equation}\label{eq1}
			\begin{split}
				\mathcal{L}=&\sum_{B}\bar{\psi}_{B}(i\gamma_{\mu}\partial^{\mu}-m_{B}+g_{\sigma B}\sigma-g_{\omega B}\gamma_{\mu}\omega^{\mu}-\tfrac12 g_{\rho B}\gamma_{\mu}\boldsymbol{\tau}.\boldsymbol{\rho}^{\mu})\psi_{B}\\
				&+\tfrac12(\partial_{\mu}\sigma\partial^{\mu}\sigma-m_{\sigma}^{2}\sigma^{2})-\tfrac14\omega_{\mu\nu}\omega^{\mu\nu}+\tfrac12 m_{\omega}^{2}\omega_{\mu}\omega^{\mu}\\
				&-\tfrac14 \boldsymbol{\rho}_{\mu\nu}.\boldsymbol{\rho}^{\mu\nu}+\tfrac12 m_{\rho}^{2}\boldsymbol{\rho}_{\mu}.\boldsymbol{\rho}^{\mu}-\tfrac13 bm_{n}(g_{\sigma}\sigma)^{3}-\tfrac14 c(g_{\sigma}\sigma)^{4}\\
				&+\sum_{\lambda}\bar{\psi}_{\lambda}(i\gamma_{\mu}\partial^{\mu}-m_{\lambda})\psi_{\lambda}.
			\end{split}
		\end{equation}
	Where $ B $ indicates the baryons and $ \lambda $ expresses leptons. $\sigma$, $\omega$, and $\rho$ show sigma, omega, and rho mesons, respectively, while $\psi_{B}$ represents baryon spinors. Also, the nucleon coupling constants of this lagrangian are expressed in Table \ref{tab1}.
	\begin{table}[h]
		\begin{center}
			\begin{tabular}[h!]{|c|c|c|c|c|}
				\hline
				\textbf{$ (g_{\sigma}/m_{\sigma})^2 $ $ fm^2 $} & 
				\textbf{$ (g_{\omega}/m_{\omega})^2 $ $ fm^2 $} &
				\textbf{$ (g_{\rho}/m_{\rho})^2 $ $ fm^2 $} &
				\textbf{$ b $} &
				\textbf{$c $} \\
				\hline
				\hline
				9.031 & 4.733 & 4.825 & 0.003305 & 0.01529 \\
				\hline
			\end{tabular}
		\end{center}
		\caption{Nucleon coupling constants of the $ \sigma-\omega-\rho $ model.}
		\label{tab1}
	\end{table}
	In this paper, we used the model for hyperon coupling constants which were introduced in article \cite{r3}. The hyperon coupling constants $ (g_{\sigma B},g_{\omega B},g_{\rho B}) $ were introduced as a ratio $(x_{\sigma},x_{\omega},x_{\rho})$ of  nucleon coupling constants $ (g_{\sigma}, g_{\omega}, g_{\rho}) $ defined as below,
	\begin{align}
		x_{\sigma} &= g_{\sigma B}/g_{\sigma},\notag \\ x_{\omega}&=g_{\omega B}/g_{\omega}, \notag \\ x_{\rho}&=g_{\rho B}/g_{\rho}.
	\end{align}
	\begin{equation}\label{eq21}
		x_{\sigma}=x_{\omega}=x_{\rho}= (\frac{m_{baryon}}{m_{nucleon}})^{\zeta}, \qquad \qquad \zeta \in \Re.
	\end{equation}
    Where $ m $ indicates the mass of particles and $ \zeta $ assumes as a free parameter.
	
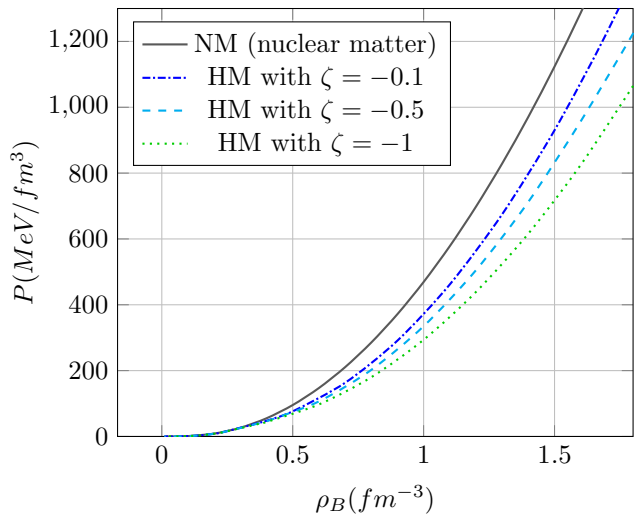
\begin{figure}[h!]
    	\begin{center}
    		\begin{tikzpicture}
    			\begin{axis}[grid,legend pos =north west,ymin=0.0001,ymax=1300,xmax=1.8,xlabel= $ \rho_{B} (fm^{-3}) $,ylabel=$  P ( MeV/ fm^{3} ) $ ]
    				\addplot[thick,smooth,gray!70!black,solid] table{0z.txt};
    				\addplot[thick,smooth,blue, densely dashdotted] table{0.1zeta.txt};
    				\addplot[thick,smooth,cyan, dashed] table{0.5zeta.txt};
    				\addplot[thick,smooth,green!80!black, dotted] table{1zeta.txt};
    				\legend{NM (nuclear matter),HM with $ \zeta=-0.1 $,HM with $ \zeta=-0.5 $,HM with $ \zeta=-1 $}
    			\end{axis}
    		\end{tikzpicture}
    	\end{center}
    	\caption{Pressure vs baryon number density for nuclear matter (NM) and hadronic matter (HM) for different values of $ \zeta $ in equilibrium conditions.}
    	\label{fig1}
    \end{figure}

	Therefore,by solving the Euler-Lagrange equations and using energy-momentum tensor, energy and pressure can be obtained as below,
	\begin{align}
		\epsilon\quad &=\quad \tfrac13 b m_{n}(g_{\sigma}\sigma)^{3}+\tfrac14 c(g_{\sigma}\sigma)^{4}+\tfrac12 m_{\sigma}^{2}\sigma^{2}+\tfrac12 m_{\omega}^{2}\omega_{0}^{2}+\tfrac12 m_{\rho}^{2}\rho_{0 3}^{2}\notag\\
		&\quad\qquad+\sum_{B}\frac{2J_{B}+1}{2\pi^{2}}\int_{0}^{k_{B}}\sqrt{k^{2}+(m_{B}-g_{\sigma B}\sigma)^{2}}k^{2}\mathrm{d}k\notag\\
		\label{neq3}&\quad\qquad+\sum_{\lambda}\frac{1}{\pi^{2}}\int_{0}^{k_{\lambda}}\sqrt{k^{2}+m_{\lambda}^{2}}k^{2}\mathrm{d}k\\
		p\quad &=\quad -\tfrac13 b m_{n}(g_{\sigma}\sigma)^{3}-\tfrac14 c(g_{\sigma}\sigma)^{4}-\tfrac12 m_{\sigma}^{2}\sigma^{2}+\tfrac12 m_{\omega}^{2}\omega_{0}^{2}+\tfrac12 m_{\rho}^{2}\rho_{0 3}^{2}\notag\\
		&\quad\qquad+\tfrac13 \sum_{B}\frac{2J_{B}+1}{2\pi^{2}}\int_{0}^{k_{B}}k^{4}\mathrm{d}k/\sqrt{k^{2}+(m_{B}-g_{\sigma B}\sigma)^{2}}\notag\\
		\label{neq4}&\quad\qquad+\tfrac13\sum_{\lambda}\frac{1}{\pi^{2}}\int_{0}^{k_{\lambda}}\frac{k^{4}}{\sqrt{k^{2}+m_{\lambda}^{2}}}\mathrm{d}k.
	\end{align}
    Where $ k_{B} $ and $ k_{\lambda} $ are the Fermi momentum of baryon $ B $ and lepton $ \lambda $. To be more clear, in Figure \ref{fig1}, we illustrate the EOS of the hadronic phase for different values of $ \zeta $. In this study, we consider the components of hadronic matter (HM) as neutrons, protons, sigmas, lambdas, electrons, and muons. Due to similarity in calculation and simplicity, other kinds of hyperons (Xi and Omega) are not considered. Also, the EOS of nuclear matter (NM) consists of neutrons, protons, and electrons are depicted (dashed line) in Figure \ref{fig1}. As we expected including more spices in Hadronic matter in addition to baryons makes the EOS softer.
	\subsection{Quark phase}
	\subsubsection{The MIT bag model}
	We first review the MIT bag model \cite{r06}.  The thermodynamic potential of quarks with flavor $ f = u, d, s $ at zero temperature can be written as below,
	\begin{align}
		\Omega_{f}(\mu_{f})&=\frac{-1}{4\pi^{2}}\left[\mu_{f}\left(\mu_{f}^{2}-\tfrac52 m_{f}^{2}\right)\sqrt{\mu_{f}^{2}-m_{f}^{2}}+\tfrac32m_{f}^{4}\ln\left(\frac{\mu_{f}+\sqrt{\mu_{f}^{2}-m_{f}^{2}}}{m_{f}}\right)\right]\notag\\&\quad+\frac{\alpha_{c}}{2\pi^{3}}\left[3\left(\mu_{f}\sqrt{\mu_{f}^{2}-m_{f}^{2}}-m_{f}^{2}\ln\left(\frac{\mu_{f}+\sqrt{\mu_{f}^{2}-m_{f}^{2}}}{m_{f}}\right)\right)^{2}\right.\notag\\&\qquad-2\left(\mu_{f}^{2}-m_{f}^{2}\right)^{2}-3m_{f}^{4}\ln\left(\frac{m_{f}}{\mu_{f}}\right)\notag\\&\left.\qquad+6\ln\frac{\sigma_{ren}}{\mu_{f}}\left[ \mu_{f}m_{f}^{2}\left(\mu_{f}^{2}-m_{f}^{2}\right)^{1/2}-\mu_{f}^{4}\ln\left(\frac{\mu_{f}+\sqrt{\mu_{f}^{2}-m_{f}^{2}}}{m_{f}}\right)\right]\right]\label{eq11},
		\end{align}
	Where first line of equation \ref{eq11} describes the kinetic term and other lines express the one-gluon-exchange term proportional to the QCD fine structure constant $ \alpha_{c} $ \cite{r5}. Also $ \mu_{f} $ and $ m_{f} $ represent chemical potential and mass of quark with flavor $ f $, and the renormalization point $ \sigma_{ren} $ is equal to $ 313 (MeV) $. We can ignore the mass of $ u $ and $ d $ quarks, while we consider $ m_{s}=300, 150 (MeV) $ \cite{r6}. Also the thermodynamic potential $ \Omega $ is defined as:
	\begin{equation}
		\Omega = \sum_{f}\Omega_{f} +B
		\end{equation}
    Where B is the bag constant and can be characterized as a free parameter. Fundamentally B is defined as the difference between the energy density of the perturbative vacuum and the true vacuum \cite{e05}. So the number density, pressure, and the total energy density can be determined as below,
    \begin{equation}
    	n_{f}=-\frac{\partial \Omega}{\partial \mu_{f}}
    \end{equation}
    \begin{equation}
    	P=-\Omega
    \end{equation}
    \begin{equation}
    	\epsilon=\Omega + \sum_{f} \mu_{f} n_{f}
    \end{equation}
    
	\subsubsection{The NJL model}
	In this section, the three-flavor version of the NJL model is introduced. The most commonly Lagrangian in this model is written as follows \cite{r7},
	\begin{equation}
		L=\bar{q}(i\slashed{\partial}-\hat{m})q+L_{sym}+L_{det}
	\end{equation}
    Where $ q=(u,d,s)^{T} $ represents a quark field with three flavors, and the corresponding quark mass matrix is expressed as $ \hat{m}=diag (m_{u},m_{d},m_{s}) $. The Lagrangian includes two independent interaction terms determined by
    \begin{equation}
    	L_{sym}=G\sum_{a=0}^{8} \left[(\bar{q}\lambda_{a}q)^{2} + (\bar{q} + i\gamma_{5}\lambda_{a}q )^{2}\right]
    \end{equation}
    and 
    \begin{equation}
    	L_{det} = -K \left[det(\bar{q}(1+\gamma_{5})q)+ det(\bar{q}(1-\gamma_{5})q) \right].
    \end{equation}
    Where G and K repersent the coupling constants of the theory. $ L_{sym} $ expresses a $ U(3)_{L} \times U(3)_{R} $ symmetric 4-point interaction, where $ \lambda_{a} $, $ a=1,\dots,8 $ define the generators of $ SU(3) $. In flavor space, $ L_{det} $, corresponding to the 't Hooft interaction, is a determinant and is a maximally flavor-mixing six-point interaction. $ L_{det} $ is $SU(3)_{L} \times SU(3)_{R}$ symmetry and breaks the $ U(1) $ symmetry, while $ U(1) $ is unbroken by $ L_{sym} $ \cite{r8}. 
    
    Also the quark self-energy in the NJL model directs to the gap equation as below,
   \begin{equation}\label{m1}
   	M_{i} =m_{i}-4G\varphi_{i} +2K\varphi_{j}\varphi_{k}.
   \end{equation}
   Where $ M $ is the constituent quark mass, $ (i,j,k) $ is any permutation of $ (u,d,s) $, and $\varphi_{i} = <\bar{q}_{i}q>$ is the quark condensate parameter \cite{r6}.
   
   Divergent integrals appear in the NJL model and we need to find a way to regularize them. Different regularization methods can be used, however when it comes to thermodynamics, a (sharp or smooth) 3-momentum cut-off $ \Lambda_{c} $ is mostly recommended. In this study, we employ a sharp 3-momentum cutoff. The cut-off $ \Lambda $ is one of the five parameters of the NJL model. The other four parameters are:  the coupling constants $ K $ and $ G $, and the bare masses $ m_{u}=m_{d} $ and $ m_{s} $. These parameters are determined by five observables: masses of the pseudoscalar mesons $ \eta^{'} $, $ \eta $, $ K $, the pion mass $ m_{\pi} $, and the pion decay constant $ f_{\pi} $. In Tabel \ref{tab2}, three different parameter sets are shown. The set of RKH belongs to fits of Rehberg, Klevansky, and Hufner \cite{r7}, HK corresponds to the fits of Hatsuda and Kunihiro \cite{r71}, and LKW are the fits of Lutz, Klimt, and Weise \cite{r72}.
       \begin{table}
   	\begin{center}
   		\begin{tabular}[hi]{|p{1.8cm}|| p{1.5cm} p{1.5cm} p{1.5cm}|| p{2.5cm}|}
   			\hline
   			\quad & \textbf{RKH \cite{r7}} & \textbf{HK \cite{r71}} & \textbf{LKW \cite{r72}} & \textbf{Empirical \cite{r73}} \\
   			\hline
   			$\Lambda_{c} (MeV)$ & 602.3 & 631.4 & 750 & \quad \\
   			$ G \Lambda_{c}^{2} $ & 1.835 & 1.835 & 1.82 & \quad \\
   			$ K \Lambda_{c}^{5} $ & 12.36 & 9.29 & 8.9 & \quad \\
   			$ m_{u,d} (MeV) $ & 5.5 & 5.5 & 3.6 & 3.5 - 7.5 \\
   			$ m_{s} (MeV) $ & 140.7 & 135.7 & 87 & 110 - 210 \\
   			$ G_{v}/G $ & $\dots$ & $\dots$ & 1.1 & \quad \\
   			&&&& \\
   			$ f_{\pi}(MeV) $ & 92.4 & 93.0 & 93 & 92.4 \\
   			$ m_{\pi} (MeV) $ & 135.0 & 138 & 139 & 135.0, 139.6 \\
   			$ m_{K} (MeV) $ & 497.7 & 496 & 498 & 493.7, 497.7 \\
   			$ m_{\eta} (MeV) $ & 514.8 & 487 & 519 & 547.3 \\
   			$ m_{\eta^{'}} (MeV) $ & 957.8 & 958 & 963 & 957.8 \\
   			$ m_{\rho,\omega} (MeV) $ & $\dots$ & $\dots$ & 765 & 771.1, 782.6 \\
   			$ m_{K^{*}} (MeV) $ & $\dots$ & $\dots$ & 864 & 891.7, 896.1 \\
   			$ m_{\phi} (MeV) $ & $\dots$ & $\dots$ & 997 & 1019.5 \\
   			\hline
   			
   		\end{tabular}
   	\end{center}
   	
   	\caption{Three groups of parameters and corresponding quark and meson properties in the three-flavor NJL model.}
   	\label{tab2}
   \end{table}

   In this model, at zero temperature, the meanfield thermodynamic potential is defined as,
	\begin{align}
		\Omega (\mu_{f},\varphi_{f}) = &\sum_{f=u,d,s} \Omega_{M_{f}}(\mu_{f}) + 2G (\varphi_{u}^{2} + \varphi_{d}^{2}+ \varphi_{s}^{2})\notag\\&-4K \varphi_{u} \varphi_{d} \varphi_{s} + \Omega_{0}
	\end{align}
	Where $ \Omega_{M_{f}} $ is defined in Eq. \ref{equa1} and expresses the contribution of a gas of quasiparticles with mass $ M_{f} $. At zero temperature, $ \Omega_{M_{f}} $ takes the form 
	\begin{equation}\label{equa1}
		\Omega_{M_{f}}(\mu_{f}) = \frac{-N_{c}}{\pi^{2}}  \int_{P_{F,f}}^{\Lambda}  E_{p,f} p^{2} \mathrm{d}p - \mu_{f} n_{f}.
	\end{equation}
    Where $ n_{f} = \frac{(P_{F,f})^3}{\pi^{2}} $, $ E_{p,f} = \sqrt{p^{2} + M_{f}^{2}} $, and $ P_{F,f} = \sqrt{\mu_{f}^{2} - M_{f}^{2}} $ corresponds to number density, the on-shell energy, and the Fermi momentum of quark with flavor $ f $. Also $\Lambda$ is a sharp 3-momentum cutoff, and $ N_{c}=3 $ is the number of colors, respectively. The quark condensates can be find by minimizing $\Omega$ at the stationary points $ (\delta \Omega/\delta\varphi_{f}=0) $. So the quark condensates can be written as 
    \begin{equation}
    	\varphi _{f} = \frac{-N_{c}}{\pi^{2}} \int_{P_{F,f}}^{\Lambda} \frac{M_{f}}{E_{p,f}} p^{2}\mathrm{d}p.
    \end{equation}
    These equations $ (\varphi_{f}) $ must be solved self-consistently with Eq. \ref{m1}, creating a group of three coupled gap equations for the constituent masses. Also, $ \Omega_{0} $ can be determined easily by applying the conditions where pressure becomes zero in limits of $ \mu, T \rightarrow 0 $. After solving the self-consistent equations, other thermodynamic quantities can be determined:
     \begin{equation}
     	P = -\Omega, \qquad \qquad \epsilon = \Omega + \sum_{f} \mu_{f} n_{f}, \qquad\qquad n_{f}=-\frac{\partial \Omega}{\partial \mu_{f}}
     \end{equation}
    
    The weak decay $ (d \leftrightarrow u + e + \bar{\nu}_{e} \leftrightarrow s ) $ in the quark matter leads to the presence of electrons in the quark matter. In presence of quarks, electron's mass is negligible, so electrons are considered as a massless and non-interacting gas of fermions,
    \begin{equation}
    	P_{e}= \frac{\mu_{e}^{4}}{12 \pi^{2}}, \qquad\qquad \epsilon_{e}= \frac{\mu_{e}^{4}}{4 \pi^{2}}, \qquad\qquad n_{e}=\frac{\mu_{e}^{3}}{3 \pi^{2}}.
    \end{equation}
    and consequently,
    \begin{equation}
     P_{tot} = P + P_{e}\qquad\qquad \epsilon_{tot}=\epsilon+\epsilon_{e}.	
    \end{equation}
 
    Also, in the beta-stable matter, chemical potentials of quarks and electrons could be expressed,  
    \begin{equation}
    	\mu_{d}=\mu_{s}, \qquad\qquad \mu_{d}= \mu_{u}+\mu_{e},
    \end{equation}

    The charge neutrality condition in quark matter leads to 
    \begin{equation}
    	0=\frac23 n_{u}-\frac13 (n_{d}+n_{s})-n_{e},
    \end{equation}
    and the baryon number density is 
    \begin{equation}
    	\rho_{B}=\frac13 (n_{u}+n_{d}+n_{s}).
    \end{equation} 
    
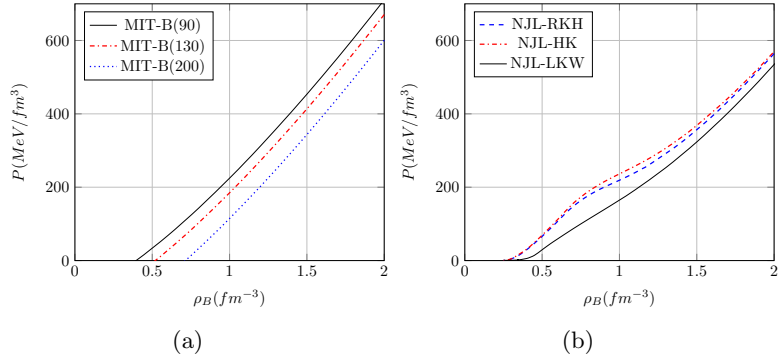
\begin{figure}[h!]
    	\centering
    	\begin{subfigure}[b]{0.4\textwidth}
    		\centering
    		\begin{tikzpicture}[scale = 0.6]
    			\begin{axis}[grid,legend pos =north west,ymin=0.0,ymax=700,xmin=0.0,xmax=2.0,xlabel= $ \rho_{B} (fm^{-3}) $,ylabel=$  P ( MeV/ fm^{3} ) $ ]
    				\addplot[smooth,black] table{P.MIT.B90.txt};
    				\addplot[thick,smooth,red, dashdotted] table{P.MIT.B130.txt};
    				\addplot[thick,smooth,blue, dotted] table{P.MIT.B200.txt};
    				
    				\legend{MIT-B(90),MIT-B(130),MIT-B(200)}
    			\end{axis}
    		\end{tikzpicture}
    	\caption{ }
    	\label{fig2a}
    	\end{subfigure}
    	\hspace*{0.1 cm}
    	\begin{subfigure}[b]{0.4\textwidth}
    		\centering
    		\begin{tikzpicture}[scale = 0.6]
    			\begin{axis}[grid,legend pos =north west,ymin=0.0,ymax=700,xmin=0.0,xmax=2.0,xlabel= $ \rho_{B} (fm^{-3}) $,ylabel=$  P ( MeV/ fm^{3} ) $ ]
    				
    				\addplot[thick,smooth,blue,dashed] table{P.NJL.RKH.txt};
    				\addplot[thick,smooth,red,dashdotted] table{P.NJL.HK.txt};
    				\addplot[smooth,black] table{P.NJL.LKW.txt};
    				\legend{NJL-RKH,NJL-HK,NJL-LKW}
    			\end{axis}
    		\end{tikzpicture}
    	\caption{ }
    	\label{fig2b}
    	\end{subfigure}
    	\caption{Pressure vs baryon number density for quark matter in equilibrium conditions; \subref{fig2a}: the MIT bag model with $ B=90 MeV fm^{-3} $, $ B=130 MeV fm^{-3} $, $ B=200 MeV fm^{-3} $ and $ m_{s} = 150 MeV $, \subref{fig2b}: the NJL model with RKH, HK, LKW parameters set.}
    	\label{fig2}
    \end{figure}

    To be more clear, we illustrate the EOS of quark matter by employing different models in Fig \ref{fig2}.
	
	\section{Results}\label{sec3}
	\subsection{Phase transition and the EOS}
	In this article, we study two types of phase transition namely Maxwell construction and Gibbs construction, which are well-known in Hadron-Quark phase transition studies. According to Maxwell's construction, phase transition arises when the baryon chemical potential and pressure of each of the individual charge neutral phases become equal i.e \cite{r9},
	\begin{equation}
		P_{H}=P_{Q} \qquad\qquad and \qquad\qquad \mu_{B}^{H}=\mu_{B}^{Q}.
	\end{equation}
    Where H and Q represent hadronic and quark phases, respectively. The other independent chemical potential $ (\mu_{e}) $ is no longer continuous during this type of phase transition$ (\mu_{e}^{H} \not=\mu_{e}^{Q}) $. The chemical potential of the electrons have a jump at the phase transition region. In addition, each phase is individually in chemical equilibrium and charge neutrality. Also, in this case, the mixed phase is no longer available.
    
    By contrast, according to Gibbs's criteria, both independent chemical potentials $ (\mu_{e}, \mu_{B}) $ are continuous. Therefore, the pressure, the baryon chemical potential, and the electron chemical potential are equal for each phase i.e,
    \begin{equation}
    	P_{H}=P_{Q}=P_{MP} \qquad , \qquad \mu_{B}^{H}=\mu_{B}^{Q} \qquad and \qquad \mu_{e}^{H} =\mu_{e}^{Q}.
    \end{equation}
    Where $ P_{MP} $ represents pressure in the mixed phase. The mixed phase is characterized by the $ \chi $ parameter, being   between zero and one. basicly, the $ \chi $ parameter deteminde three different phases as below,
    \begin{itemize}
    	\item Hadronic phase $ \qquad\qquad \chi\leq0 $,
    	\item Mixed phase $ \qquad\qquad 0<\chi<1 $,
    	\item Pure quark phase $ \qquad\qquad \chi\geq1 $.
    \end{itemize}
    In mixed phase, we apply the global conservation \cite{e06}. Therefore the baryon number density and the charge neutrality can be written as a function of $ \chi $,
    \begin{equation}
    	(1-\chi)\rho_{H}+\chi\rho_{Q}=\rho_{B},
    \end{equation}
    \begin{equation}
    	(1-\chi)q_{H}+\chi q_{Q}=0.
    \end{equation}
    Where the subscripts Q and H express quark and confined hadronic phases.
    \subsubsection{The MIT bag model (Maxwell criteria)}
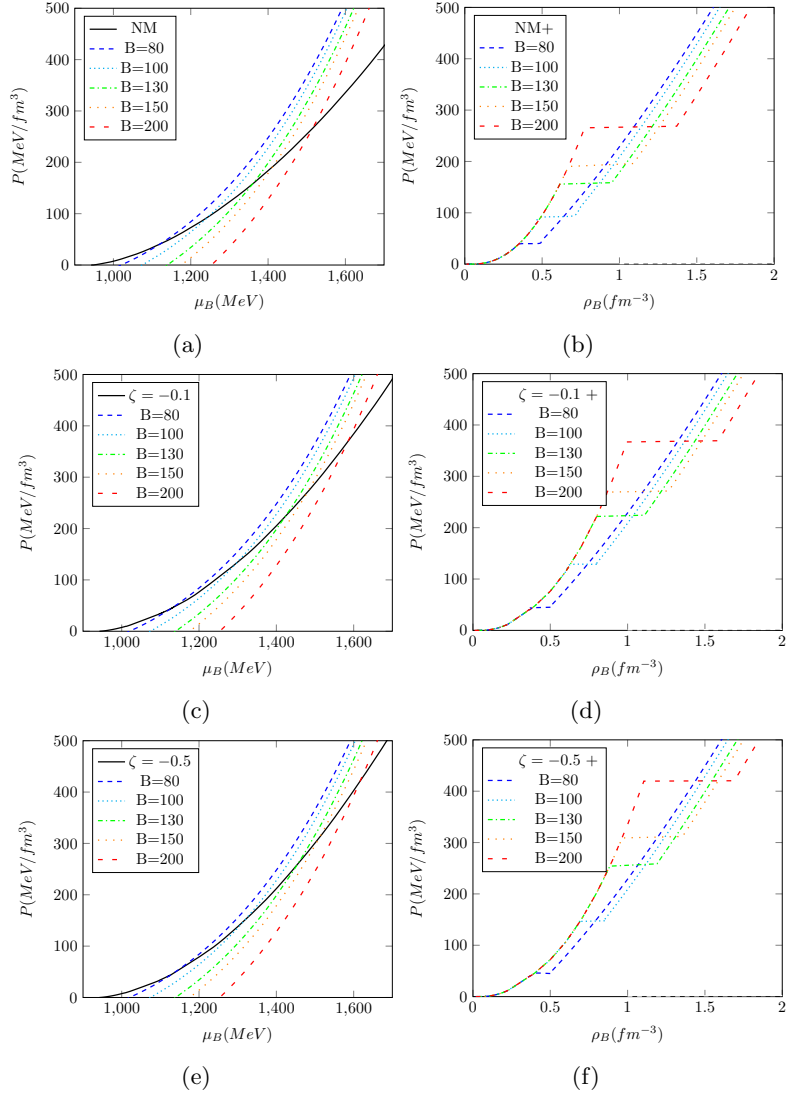
\begin{figure}[h!]
        	\centering
    	\begin{subfigure}[b]{0.4\textwidth}
    		\centering
    			\begin{tikzpicture}[scale = 0.6]
    				\begin{axis}[legend pos =north west,ymin=0.0,ymax=500,xmin=900,xmax=1700,xlabel= $ \mu_{B} (MeV) $,ylabel=$  P ( MeV/ fm^{3} ) $ ]
    					\addplot[thick,smooth,black] table{P.MU.0.txt};
    					\addplot[thick,smooth,blue,dashed] table{P.MU.B80.txt};
    					\addplot[thick,smooth,cyan,dotted] table{P.MU.B100.txt};
    					\addplot[thick,smooth,green,dashdotted] table{P.MU.B130.txt};
    					\addplot[thick,smooth,orange,loosely dotted] table{P.MU.B150.txt};
    					\addplot[thick,smooth,red,loosely dashed] table{P.MU.B200.txt};
    					\legend{NM,B=80,B=100,B=130,B=150,B=200}
    				\end{axis}
    			\end{tikzpicture}
    		\caption{ }
    		\label{fig3a}
    	\end{subfigure}
    \hspace*{0.1 cm}
    \begin{subfigure}[b]{0.4\textwidth}
    	\centering
    	\begin{tikzpicture}[scale = 0.6]
    		\begin{axis}[legend pos =north west,ymin=0.0,ymax=500,xmin=0,xmax=2,xlabel= $ \rho_{B} (fm^{-3}) $ ,ylabel=$  P ( MeV/ fm^{3} ) $ ]
    			\addplot[ thick,smooth,white,dashed] table{12.txt};
    			\addplot[ thick,smooth,blue,dashed] table{npb80.txt};
    			\addplot[ thick,smooth,cyan,dotted] table{npb100.txt};
    			\addplot[ thick,green,dashdotted] table{npb130.txt};
    			\addplot[ thick,orange,loosely dotted] table{npb150.txt};
    			\addplot[ thick,red,loosely dashed] table{npb200.txt};

    			\legend{NM+,B=80,B=100,B=130,B=150,B=200}
    		\end{axis}
    	\end{tikzpicture}
    \caption{ }
    \label{fig3b}
    \end{subfigure}
    \hspace*{0.1 cm}
        \begin{subfigure}[b]{0.4\textwidth}
        	\centering
        	\begin{tikzpicture}[scale = 0.6]
        		\begin{axis}[legend pos =north west,ymin=0.0,ymax=500,xmin=900,xmax=1700,xlabel= $ \mu_{B} (MeV) $,ylabel=$  P ( MeV/ fm^{3} ) $ ]
        			\addplot[ thick,smooth,black] table{P.MU.0.1.txt};
        			\addplot[ thick,smooth,blue,dashed] table{P.MU.B80.txt};
        			\addplot[ thick,smooth,cyan,dotted] table{P.MU.B100.txt};
        			\addplot[ thick,smooth,green,dashdotted] table{P.MU.B130.txt};
        			\addplot[ thick,smooth,orange,loosely dotted] table{P.MU.B150.txt};
        			\addplot[ thick,smooth,red,loosely dashed] table{P.MU.B200.txt};
        			\legend{$ \zeta=-0.1 $,B=80,B=100,B=130,B=150,B=200}
        		\end{axis}
        	\end{tikzpicture}
        \caption{ }
        \label{fig3c}
        \end{subfigure}
        \hspace*{0.1 cm}
        \begin{subfigure}[b]{0.4\textwidth}
        	\centering
        	\begin{tikzpicture}[scale = 0.6]
        		\begin{axis}[legend pos =north west,ymin=0.0,ymax=500,xmin=0,xmax=2,xlabel= $ \rho_{B} (fm^{-3}) $ ,ylabel=$  P ( MeV/ fm^{3} ) $ ]
        			\addplot[ thick,smooth,white,dashed] table{12.txt};
        			\addplot[ thick,smooth,blue,dashed] table{0.1b80.txt};
        			\addplot[ thick,smooth,cyan,dotted] table{0.1b100.txt};
        			\addplot[ thick,green,dashdotted] table{0.1b130.txt};
        			\addplot[ thick,orange,loosely dotted] table{0.1b150.txt};
        			\addplot[ thick,red,loosely dashed] table{0.1b200.txt};

        			\legend{$ \zeta=-0.1 $ +,B=80,B=100,B=130,B=150,B=200}
        		\end{axis}
        	\end{tikzpicture}
        \caption{ }
        \label{fig3d}
        \end{subfigure}
        \hspace*{0.1 cm}
        \begin{subfigure}[b]{0.4\textwidth}
        	\centering
        	\begin{tikzpicture}[scale = 0.6]
        		\begin{axis}[legend pos =north west,ymin=0.0,ymax=500,xmin=900,xmax=1700,xlabel= $ \mu_{B} (MeV) $,ylabel=$  P ( MeV/ fm^{3} ) $ ]
        			\addplot[ thick,smooth,black] table{P.MU.0.5.txt};
        			\addplot[ thick,smooth,blue,dashed] table{P.MU.B80.txt};
        			\addplot[ thick,smooth,cyan,dotted] table{P.MU.B100.txt};
        			\addplot[ thick,smooth,green,dashdotted] table{P.MU.B130.txt};
        			\addplot[ thick,smooth,orange,loosely dotted] table{P.MU.B150.txt};
        			\addplot[ thick,smooth,red,loosely dashed] table{P.MU.B200.txt};
        			\legend{$ \zeta=-0.5 $,B=80,B=100,B=130,B=150,B=200}
        		\end{axis}
        	\end{tikzpicture}
        \caption{ }
        \label{fig3e}
        \end{subfigure}
        \hspace*{0.1 cm}
        \begin{subfigure}[b]{0.4\textwidth}
        	\centering
        	\begin{tikzpicture}[scale = 0.6]
        		\begin{axis}[legend pos =north west,ymin=0.0,ymax=500,xmin=0,xmax=2,xlabel= $ \rho_{B} (fm^{-3}) $ ,ylabel=$  P ( MeV/ fm^{3} ) $ ]
        			\addplot[ thick,smooth,white,dashed] table{12.txt};
        			\addplot[ thick,smooth,blue,dashed] table{0.5b80.txt};
        			\addplot[ thick,smooth,cyan,dotted] table{0.5b100.txt};
        			\addplot[thick,green,dashdotted] table{0.5b130.txt};
        			\addplot[thick,orange,loosely dotted] table{0.5b150.txt};
        			\addplot[thick,red,loosely dashed] table{0.5b200.txt};

        			\legend{$ \zeta=-0.5 $ +,B=80,B=100,B=130,B=150,B=200}
        		\end{axis}
        	
        	\end{tikzpicture}
        \caption{ }
        \label{fig3f}
        \end{subfigure}
        \caption{Left: Pressure vs. baryon chemical potential for the MIT bag model with different values of B combined with nuclear matter (NM) and hadron matter with $ \zeta=-0.1 $ and $ \zeta=-0.5 $, respectively. Right:The corresponding hadron-quark hybrid EoS within the Maxwell phase transition.}
        \label{fig3}
    \end{figure}

    We first discuss the equation of state (EOS) of hybrid matter within the Maxwell phase transition. For hadronic phases, we use two kinds of the equation of state (EOS). First is Nuclear matter (NM) consisting of neutrons, protons, and electrons. And second one (characterized by the $ \zeta $ parameter) consists of neutrons, protons, sigmas, lamdas, muons, and electrons. Also, the quark phase is described by the MIT bag model with different values of B (the bag constant). In our calculation, we consider $ 80 MeV fm^{-3}\leq B \leq 200 MeV fm^{-3}$. For very small $ B $, the phase transition happens at a point that is close to the saturation density, or at that point, there is no possibility of hyperon presence, so we consider $ B\geq 80 MeV fm^{-3} $. The results are shown in Figs \ref{fig3}.  

    As can be seen in Figure \ref{fig3}, panels \subref{fig3a} and \subref{fig3b} show the results for nuclear matter transition to Quark matter using MIT bag model, while panels \subref{fig3c} and \subref{fig3d} ( \subref{fig3e} and \subref{fig3f}) show the results for hadronic matter including hyperons with $ \zeta=-0.1 (-0.5) $. Also, with the increase of B, the starting point of the phase transition occurs at higher densities, and the density difference between the hadronic and quark phases increases. In addition, as the value of $ \zeta $ decreases, the phase transition point increases again and occurs at higher densities $ (\rho = 7-8 \rho_{0} ) $.
    \subsubsection{The MIT bag model (Gibbs's criteria)}
\begin{figure}[h!]
    	\centering
    \begin{subfigure}[b]{0.4\textwidth}
    	\centering
    	\begin{tikzpicture}[scale = 0.6]

    		\begin{axis}[legend pos =north west,ymin=0.0,ymax=600,xmin=0,xmax=2.1,xlabel= $ \rho_{B} (fm^{-3}) $,ylabel=$  P ( MeV/ fm^{3} ) $ ,axis on top,legend style={legend cell align=right,legend plot pos=right}]
    			
    			\addplot+[mark=none,
    			domain=0.1:0.5,
    			samples=100,
    			pattern=grid,
    			draw=blue,
    			area legend,
    			pattern color=blue]table{N-PB230S3001.txt} \closedcycle;
    			\addlegendentry{Pure hadronic phase}
    			
    			\addplot+[mark=none,
    			domain=0.1:0.5,
    			samples=100,
    			pattern=dots,
    			draw=black,
    			area legend,
    			pattern color=black]table{N-PB230S3002.txt} \closedcycle;
    			\addlegendentry{Mixed phase}
    			
    			\addplot+[mark=none,
    			domain=0.1:0.5,
    			samples=100,
    			pattern=crosshatch,
    			draw=orange,
    			area legend,
    			pattern color=orange]table{N-PB230S3003.txt} \closedcycle;
    			\addlegendentry{Pure quark phase}
    			
    			\addplot[very thick,smooth,black,domain=0:2,samples=100] table{N-PB230S300.txt};
    			\addlegendentry{Pressure}
    			
    		\end{axis}
    	\end{tikzpicture}
    \caption{ }
    \label{fig4a}
    \end{subfigure}
    \hspace*{0.1 cm}
    \begin{subfigure}[b]{0.4\textwidth}
    	\centering
    	\begin{tikzpicture}[scale = 0.6]
    		\tikzset{
    			hatch distance/.store in=\hatchdistance,
    			hatch distance=10pt,
    			hatch thickness/.store in=\hatchthickness,
    			hatch thickness=1pt
    		}
    		
    		\makeatletter
    		\pgfdeclarepatternformonly[\hatchdistance,\hatchthickness]{flexible hatch}
    		{\pgfqpoint{0pt}{0pt}}
    		{\pgfqpoint{\hatchdistance}{\hatchdistance}}
    		{\pgfpoint{\hatchdistance-1pt}{\hatchdistance-1pt}}%
    		{
    			\pgfsetcolor{\tikz@pattern@color}
    			\pgfsetlinewidth{\hatchthickness}
    			\pgfpathmoveto{\pgfqpoint{0pt}{0pt}}
    			\pgfpathlineto{\pgfqpoint{\hatchdistance}{\hatchdistance}}
    			\pgfusepath{stroke}
    		}
    		
    		\begin{axis}[legend pos =north west,ymin=0.0,ymax=600,xmin=0,xmax=2,xlabel= $ \rho_{B} (fm^{-3}) $,ylabel=$  P ( MeV/ fm^{3} ) $ ,axis on top,legend style={legend cell align=right,legend plot pos=right}]
    			\addplot+[mark=none,
    			domain=0.1:0.5,
    			samples=100,
    			pattern=grid,
    			draw=blue,
    			area legend,
    			pattern color=blue]table{N-PB210S3001.txt} \closedcycle;
    			\addlegendentry{Pure hadronic phase}
    			
    			\addplot+[mark=none,
    			domain=0.1:0.5,
    			samples=100,
    			pattern=dots,
    			draw=black,
    			area legend,
    			pattern color=black]table{N-PB210S3002.txt} \closedcycle;
    			\addlegendentry{Mixed phase}
    			
    			\addplot+[mark=none,
    			domain=0.1:0.5,
    			samples=100,
    			pattern=crosshatch,
    			draw=orange,
    			area legend,
    			pattern color=orange]table{N-PB210S3003.txt} \closedcycle;
    			\addlegendentry{Pure quark phase}
    			
    			\addplot[very thick,smooth,black,domain=0:2,samples=100] table{N-PB210S300.txt};
    			\addlegendentry{Pressure}
    			
    		\end{axis}
    	\end{tikzpicture}
    	\caption{ }
    	\label{fig4b}
    \end{subfigure}
    \hspace*{0.1 cm}
    \begin{subfigure}[b]{0.4\textwidth}
    	\centering
    	\begin{tikzpicture}[scale = 0.6]

    		\begin{axis}[legend pos =north west,ymin=0.0,ymax=600,xmin=0,xmax=2.0,xlabel= $ \rho_{B} (fm^{-3}) $,ylabel=$  P ( MeV/ fm^{3} ) $ ,axis on top,legend style={legend cell align=right,legend plot pos=right}]

    			\addplot+[mark=none,
    			domain=0.1:0.5,
    			samples=100,
    			pattern=grid,
    			draw=blue,
    			area legend,
    			pattern color=blue]table{Z0.1B1601.txt} \closedcycle;
    			\addlegendentry{Pure hadronic phase}
    			
    			\addplot+[mark=none,
    			domain=0.1:0.5,
    			samples=100,
    			pattern=dots,
    			draw=black,
    			area legend,
    			pattern color=black]table{Z0.1B1602.txt} \closedcycle;
    			\addlegendentry{Mixed phase}
    			
    			\addplot+[mark=none,
    			domain=0.1:0.5,
    			samples=100,
    			pattern=crosshatch,
    			draw=orange,
    			area legend,
    			pattern color=orange]table{Z0.1B1603.txt} \closedcycle;
    			\addlegendentry{Pure quark phase}
    			
    			\addplot[very thick,smooth,black,domain=0:2,samples=100] table{Z0.1B160.txt};
    			\addlegendentry{Pressure}
    			
    		\end{axis}
    	\end{tikzpicture}
    \caption{ }
    \label{fig4c}
    \end{subfigure}
    \hspace*{0.1 cm}
\begin{subfigure}[b]{0.4\textwidth}
	\centering
	\begin{tikzpicture}[scale = 0.6]

		\begin{axis}[legend pos =north west,ymin=0.0,ymax=600,xmin=0,xmax=2.0,xlabel= $ \rho_{B} (fm^{-3}) $,ylabel=$  P ( MeV/ fm^{3} ) $ ,axis on top,legend style={legend cell align=right,legend plot pos=right}]

			\addplot+[mark=none,
			domain=0.1:0.5,
			samples=100,
			pattern=grid,
			draw=blue,
			area legend,
			pattern color=blue]table{Z0.5B1401.txt} \closedcycle;
			\addlegendentry{Pure hadronic phase}
			
			\addplot+[mark=none,
			domain=0.1:0.5,
			samples=100,
			pattern=dots,
			draw=black,
			area legend,
			pattern color=black]table{Z0.5B1402.txt} \closedcycle;
			\addlegendentry{Mixed phase}
			
			\addplot+[mark=none,
			domain=0.1:0.5,
			samples=100,
			pattern=crosshatch,
			draw=orange,
			area legend,
			pattern color=orange]table{Z0.5B1403.txt} \closedcycle;
			\addlegendentry{Pure quark phase}
			
			\addplot[very thick,smooth,black,domain=0:2,samples=100] table{Z0.5B140.txt};
			\addlegendentry{Pressure}
			
		\end{axis}
	\end{tikzpicture}
\caption{ }
\label{fig4d}
\end{subfigure}

\caption{Pressure vs baryon number density for hybrid matter within the Gibbs phase transition. \subref{fig4a}: Nuclear matter combined with the MIT bag model $( B=230 MeV fm^{-3} )$; \subref{fig4b}: Nuclear matter combined with the MIT bag model $( B=210 MeV fm^{-3})$; \subref{fig4c}: Hadron matter $ (\zeta=-0.1) $ combined with the MIT bag model $( B=160 MeV fm^{-3} )$; \subref{fig4d}: Hadron matter $ (\zeta=-0.5) $ combined with the MIT bag model $( B=140 MeV fm^{-3} )$.}
\label{fig4}
    \end{figure}
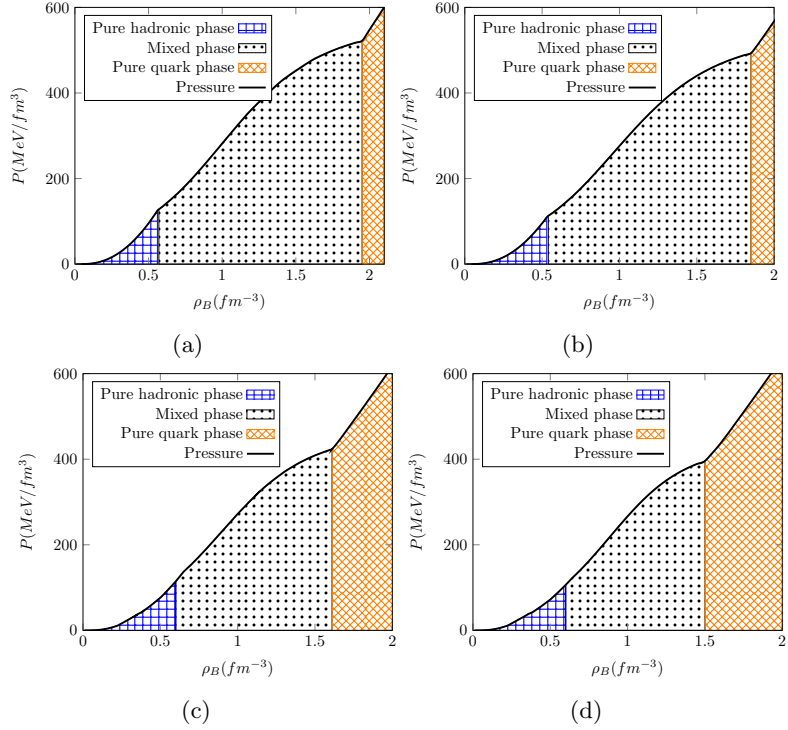

    In this section, the results related to phase transition with Gibbs's criteria are presented. Here, as before, two types of equation of state are used to describe the hadronic phase: nuclear matter (NM) and hadronic matter (HM) with different values for $ \zeta $. Also, to describe the quark phase, the MIT bag model with different values of B has been used. The results are shown in Fig. \ref{fig4}.

      As mentioned earlier, in the phase transition under Gibbs conditions, there is a mixed phase region where both the hadronic phase and quark phase coexist under a specific ratio of $ \chi $ in each density. As shown in Figure \ref{fig4}, the length of the mixed phase is directly related to the value of B and increases with the increasing of B. Also, the starting density of the mixed phase does not depend significantly on the equation of states used and occurs around the $ \rho=0.5 (fm^{-3}) $.
     \subsubsection{The NJL model (Maxwell criteria)}
\begin{figure}[h!]
     	\centering
     	\begin{subfigure}[b]{0.4\textwidth}
     		\centering
     		\begin{tikzpicture}[scale = 0.6]
     			\begin{axis}[legend pos =north west,ymin=0.0,ymax=500,xmin=900,xmax=1700,xlabel= $ \mu_{B} (MeV) $,ylabel=$  P ( MeV/ fm^{3} ) $ ]
     				\addplot[very thick,smooth,black] table{P.MU.RKH.txt};
     				\addplot[very thick,smooth,blue,dashed] table{P.MU.0.txt};
     				\addplot[very thick,smooth,green,loosely dotted] table{P.MU.0.1.txt};
     				\addplot[very thick,smooth,red,loosely dashdotted] table{P.MU.0.5.txt};
     				
     				\legend{RKH,NM,$ \zeta=-0.1 $,$ \zeta=-0.5 $}
     			\end{axis}
     		\end{tikzpicture}
     	\caption{ }
     	\label{fig5a}
     	\end{subfigure}
     	\hspace*{0.1 cm}
     	\begin{subfigure}[b]{0.4\textwidth}
     		\centering
     		\begin{tikzpicture}[scale = 0.6]
     			\begin{axis}[legend pos =north west,ymin=0.0,ymax=500,xmin=0,xmax=2,xlabel= $ \rho_{B} (fm^{-3}) $ ,ylabel=$  P ( MeV/ fm^{3} ) $ ]
     				\addplot[very thick,smooth,white,dashed] table{12.txt};
     				\addplot[very thick,smooth,blue,dashed] table{n-p-rkh.txt};
     				\addplot[very thick,smooth,green,loosely dotted] table{z0.1-rkh.txt};
     				\addplot[very thick,smooth,red,loosely dashdotted] table{z0.5-rkh.txt};
     				
     				\legend{RKH+,NM,$ \zeta=-0.1 $,$ \zeta=-0.5 $}
     			\end{axis}
     		\end{tikzpicture}
     	\caption{ }
     	\label{fig5b}
     	\end{subfigure}
     	\hspace*{0.1 cm}
     	\begin{subfigure}[b]{0.4\textwidth}
     		\centering
     		\begin{tikzpicture}[scale = 0.6]
     			\begin{axis}[legend pos =north west,ymin=30,ymax=500,xmin=900,xmax=1700,xlabel= $ \mu_{B} (MeV) $,ylabel=$  P ( MeV/ fm^{3} ) $ ]
     				\addplot[very thick,smooth,black] table{P.MU.LKW.txt};
     				\addplot[very thick,smooth,blue,dashed] table{P.MU.0.txt};
     				\addplot[very thick,smooth,green,loosely dotted] table{P.MU.0.1.txt};
     				\addplot[very thick,smooth,red,loosely dashdotted] table{P.MU.0.5.txt};
     				
     				\legend{LKW,NM,$ \zeta=-0.1 $,$ \zeta=-0.5 $}
     			\end{axis}
     		\end{tikzpicture}
     	\caption{ }
     	\label{fig5c}
     	\end{subfigure}
     	\hspace*{0.1 cm}
     	\begin{subfigure}[b]{0.4\textwidth}
     		\centering
     		\begin{tikzpicture}[scale = 0.6]
     			\begin{axis}[legend pos =north west,ymin=0.0,ymax=500,xmin=0,xmax=2,xlabel= $ \rho_{B} (fm^{-3}) $ ,ylabel=$  P ( MeV/ fm^{3} ) $ ]
     				\addplot[very thick,smooth,white,dashed] table{12.txt};
     				\addplot[very thick,smooth,blue,dashed] table{n-p-lkw.txt};
     				\addplot[very thick,smooth,green,loosely dotted] table{z0.1-lkw.txt};
     				\addplot[very thick,smooth,red,loosely dashdotted] table{z0.5-lkw.txt};
     				
     				\legend{LKW+,NM,$ \zeta=-0.1 $,$ \zeta=-0.5 $}
     			\end{axis}
     		\end{tikzpicture}
     	\caption{ }
     	\label{fig5d}
     	\end{subfigure}
     	\hspace*{0.1 cm}
     	\begin{subfigure}[b]{0.4\textwidth}
     		\centering
     		\begin{tikzpicture}[scale = 0.6]
     			\begin{axis}[legend pos =north west,ymin=0.0,ymax=500,xmin=900,xmax=1700,xlabel= $ \mu_{B} (MeV) $,ylabel=$  P ( MeV/ fm^{3} ) $ ]
     				\addplot[very thick,smooth,black] table{P.MU.HK.txt};
     				\addplot[very thick,smooth,blue,dashed] table{P.MU.0.txt};
     				\addplot[very thick,smooth,green,loosely dotted] table{P.MU.0.1.txt};
     				\addplot[very thick,smooth,red,loosely dashdotted] table{P.MU.0.5.txt};
     				
     				\legend{HK,NM,$ \zeta=-0.1 $,$ \zeta=-0.5 $}
     			\end{axis}
     		\end{tikzpicture}
     	\caption{ }
     	\label{fig5e}
     	\end{subfigure}
     	    \hspace*{0.1 cm}
     	    \begin{subfigure}[b]{0.4\textwidth}
     	    	\centering
     	    	\begin{tikzpicture}[scale = 0.6]
     	    		\begin{axis}[legend pos =north west,ymin=0.0,ymax=500,xmin=0,xmax=2,xlabel= $ \rho_{B} (fm^{-3}) $ ,ylabel=$  P ( MeV/ fm^{3} ) $ ]
     	    			\addplot[very thick,smooth,white,dashed] table{12.txt};
     	    			\addplot[very thick,smooth,blue,dashed] table{n-p-hk.txt};
     	    			\addplot[very thick,smooth,green,loosely dotted] table{z0.1-hk.txt};
     	    			\addplot[very thick,smooth,red,loosely dashdotted] table{z0.5-hk.txt};
     	    			
     	    			\legend{HK+,NM,$ \zeta=-0.1 $,$ \zeta=-0.5 $}
     	    		\end{axis}
     	    	\end{tikzpicture}
         	\caption{ }
         	\label{fig5f}
     	\end{subfigure}
     	\caption{Left: Pressure vs. baryon chemical potential for the NJL model with different parameter sets (RKH, LKW, HK) combined with nuclear matter (NM) and hadron matter with $ \zeta=-0.1 $ and $ \zeta=-0.5 $, respectively. Right:The corresponding hadron-quark hybrid EoS within the Maxwell phase transition.}
     	\label{fig5}
     \end{figure}
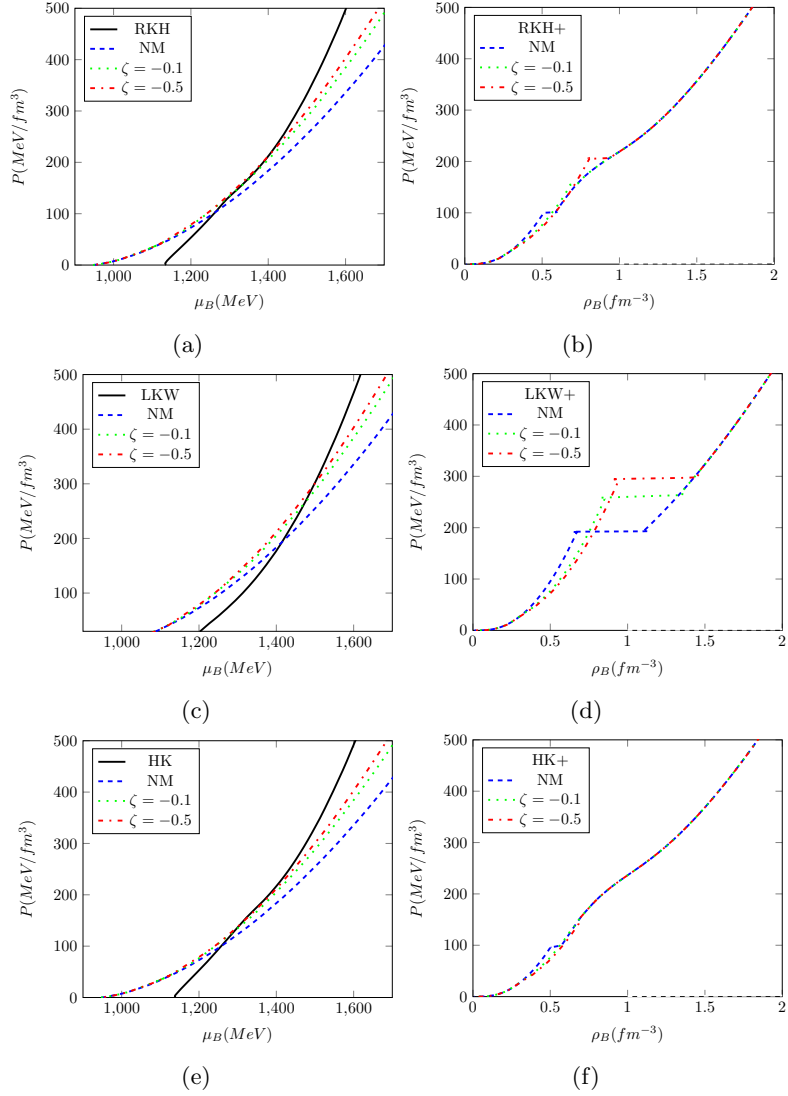

     In this section, the NJL model with different parameter sets (RKH, LKW, HK) is used to describe quark matter, and the equation of state of hadronic matter is the same as before. Also, Maxwell's phase transition conditions have been used in the investigation of this hybrid matter composition. The graphs of pressure vs. baryon chemical potential and pressure vs. baryon number density of this hybrid matter are presented in Fig. \ref{fig5}.
	
	As we can see in Fig. \ref{fig5}, the phase transition has happened for all cases. In addition,  the slope of the quark phase diagram and the hadronic phase diagram are close to each other at the phase transition point, shown in figures \ref{fig5a} and \ref{fig5e}. This makes the density difference of the two phases very small in the phase transition, shown in figures \ref{fig5b} and \ref{fig5f}. For example, for the case where HK is used to describe the quark phase and $ \zeta=-0.5 $ for the hadronic phase, the density difference in the phase transition is equal to $ 0.005 fm^{-3}$. Also, with the decrease of the value of $ \zeta $, the density of the starting point of the phase transition increases. 
	
	In the phase transition under Maxwell's conditions, when the quark phase is described by the MIT model, the results of the phase transition are almost similar to those when the hadronic phase is described by the LOCV method \cite{r6}. But the main difference is related to the NJL model. As shown in Figure \ref{fig5}, the density differences between the two phases are very small, except for case \ref{fig5d}. By contrast, these density differences are significant for the LOCV method for different parameter sets.
	
	\subsubsection{The NJL model (Gibbs's criteria)}
\begin{figure}[h!]
		\centering
		\begin{subfigure}[b]{0.4\textwidth}
			\centering
			\begin{tikzpicture}[scale = 0.6]

				\begin{axis}[legend pos =north west,ymin=0.0,ymax=600,xmin=0,xmax=2.0,xlabel= $ \rho_{B} (fm^{-3}) $,ylabel=$  P ( MeV/ fm^{3} ) $ ,axis on top,legend style={legend cell align=right,legend plot pos=right}]
					
					\addplot+[mark=none,
					domain=0.1:0.5,
					samples=100,
					pattern=grid,
					draw=blue,
					area legend,
					pattern color=blue]table{N-Prkh1.txt} \closedcycle;
					\addlegendentry{Pure hadronic phase}
					
					\addplot+[mark=none,
					domain=0.1:0.5,
					samples=100,
					pattern=dots,
					draw=black,
					area legend,
					pattern color=black]table{N-Prkh2.txt} \closedcycle;
					\addlegendentry{Mixed phase}
					
					\addplot+[mark=none,
					domain=0.1:0.5,
					samples=100,
					pattern=crosshatch,
					draw=orange,
					area legend,
					pattern color=orange]table{N-Prkh3.txt} \closedcycle;
					\addlegendentry{Pure quark phase}
					
					\addplot[very thick,smooth,black,domain=0:2,samples=100] table{N-Prkh.txt};
					\addlegendentry{Pressure}
					
				\end{axis}
			\end{tikzpicture}
			\caption{NM combined with RKH}
			\label{figea}
		\end{subfigure}
	    \hspace*{0.1 cm}
	    \begin{subfigure}[b]{0.4\textwidth}
	    	\centering
	    	\begin{tikzpicture}[scale = 0.6]

	    		\begin{axis}[legend pos =north west,ymin=0.0,ymax=600,xmin=0,xmax=2.0,xlabel= $ \rho_{B} (fm^{-3}) $,ylabel=$  P ( MeV/ fm^{3} ) $ ,axis on top,legend style={legend cell align=right,legend plot pos=right}]
	    			
	    			\addplot+[mark=none,
	    			domain=0.1:0.5,
	    			samples=100,
	    			pattern=grid,
	    			draw=blue,
	    			area legend,
	    			pattern color=blue]table{0.1rkh1.txt} \closedcycle;
	    			\addlegendentry{Pure hadronic phase}
	    			
	    			\addplot+[mark=none,
	    			domain=0.1:0.5,
	    			samples=100,
	    			pattern=dots,
	    			draw=black,
	    			area legend,
	    			pattern color=black]table{0.1rkh2.txt} \closedcycle;
	    			\addlegendentry{Mixed phase}
	    			
	    			\addplot+[mark=none,
	    			domain=0.1:0.5,
	    			samples=100,
	    			pattern=crosshatch,
	    			draw=orange,
	    			area legend,
	    			pattern color=orange]table{0.1rkh3.txt} \closedcycle;
	    			\addlegendentry{Pure quark phase}
	    			
	    			\addplot[very thick,smooth,black,domain=0:2,samples=100] table{0.1rkh.txt};
	    			\addlegendentry{Pressure}
	    			
	    		\end{axis}
	    	\end{tikzpicture}
	    	\caption{$\zeta=-0.1 $ combined with RKH}
	    	\label{figeb}
	    \end{subfigure}
		\hspace*{0.01 cm}
		\begin{subfigure}[b]{0.4\textwidth}
			\centering
			\begin{tikzpicture}[scale = 0.6]

				\begin{axis}[legend pos =north west,ymin=0.0,ymax=600,xmin=0,xmax=2.0,xlabel= $ \rho_{B} (fm^{-3}) $,ylabel=$  P ( MeV/ fm^{3} ) $ ,axis on top,legend style={legend cell align=right,legend plot pos=right}]
					
					\addplot+[mark=none,
					domain=0.1:0.5,
					samples=100,
					pattern=grid,
					draw=blue,
					area legend,
					pattern color=blue]table{N-Phk1.txt} \closedcycle;
					\addlegendentry{Pure hadronic phase}
					
					\addplot+[mark=none,
					domain=0.1:0.5,
					samples=100,
					pattern=dots,
					draw=black,
					area legend,
					pattern color=black]table{N-Phk2.txt} \closedcycle;
					\addlegendentry{Mixed phase}
					
					\addplot+[mark=none,
					domain=0.1:0.5,
					samples=100,
					pattern=crosshatch,
					draw=orange,
					area legend,
					pattern color=orange]table{N-Phk3.txt} \closedcycle;
					\addlegendentry{Pure quark phase}
					
					\addplot[very thick,smooth,black,domain=0:2,samples=100] table{N-Phk.txt};
					\addlegendentry{Pressure}
					
				\end{axis}
			\end{tikzpicture}
			\caption{NM combined with HK}
			\label{figec}
		\end{subfigure}
	    \hspace*{0.1 cm}
	    \begin{subfigure}[b]{0.4\textwidth}
	    	\centering
	    	\begin{tikzpicture}[scale = 0.6]

	    		\begin{axis}[legend pos =north west,ymin=0.0,ymax=600,xmin=0,xmax=2.0,xlabel= $ \rho_{B} (fm^{-3}) $,ylabel=$  P ( MeV/ fm^{3} ) $ ,axis on top,legend style={legend cell align=right,legend plot pos=right}]
	    			
	    			\addplot+[mark=none,
	    			domain=0.1:0.5,
	    			samples=100,
	    			pattern=grid,
	    			draw=blue,
	    			area legend,
	    			pattern color=blue]table{0.5hk1.txt} \closedcycle;
	    			\addlegendentry{Pure hadronic phase}
	    			
	    			\addplot+[mark=none,
	    			domain=0.1:0.5,
	    			samples=100,
	    			pattern=dots,
	    			draw=black,
	    			area legend,
	    			pattern color=black]table{0.5hk2.txt} \closedcycle;
	    			\addlegendentry{Mixed phase}
	    			
	    			\addplot+[mark=none,
	    			domain=0.1:0.5,
	    			samples=100,
	    			pattern=crosshatch,
	    			draw=orange,
	    			area legend,
	    			pattern color=orange]table{0.5hk3.txt} \closedcycle;
	    			\addlegendentry{Pure quark phase}
	    			
	    			\addplot[very thick,smooth,black,domain=0:2,samples=100] table{0.5hk.txt};
	    			\addlegendentry{Pressure}
	    			
	    		\end{axis}
	    	\end{tikzpicture}
	    	\caption{$\zeta=-0.5 $ combined with HK}
	    	\label{figed}
	    \end{subfigure}
	    \hspace*{0.01 cm}
	    \begin{subfigure}[b]{0.4\textwidth}
	    	\centering
	    	\begin{tikzpicture}[scale = 0.6]

	    		\begin{axis}[legend pos =north west,ymin=0.0,ymax=600,xmin=0,xmax=2.0,xlabel= $ \rho_{B} (fm^{-3}) $,ylabel=$  P ( MeV/ fm^{3} ) $ ,axis on top,legend style={legend cell align=right,legend plot pos=right}]
	    			
	    			\addplot+[mark=none,
	    			domain=0.1:0.5,
	    			samples=100,
	    			pattern=grid,
	    			draw=blue,
	    			area legend,
	    			pattern color=blue]table{N-Plkw1.txt} \closedcycle;
	    			\addlegendentry{Pure hadronic phase}
	    			
	    			\addplot+[mark=none,
	    			domain=0.1:0.5,
	    			samples=100,
	    			pattern=dots,
	    			draw=black,
	    			area legend,
	    			pattern color=black]table{N-Plkw2.txt} \closedcycle;
	    			\addlegendentry{Mixed phase}
	    			
	    			\addplot+[mark=none,
	    			domain=0.1:0.5,
	    			samples=100,
	    			pattern=crosshatch,
	    			draw=orange,
	    			area legend,
	    			pattern color=orange]table{N-Plkw3.txt} \closedcycle;
	    			\addlegendentry{Pure quark phase}
	    			
	    			\addplot[very thick,smooth,black,domain=0:2,samples=100] table{N-Plkw.txt};
	    			\addlegendentry{Pressure}
	    			
	    		\end{axis}
	    	\end{tikzpicture}
	    	\caption{NM combined with LKW}
	    	\label{figee}
	    \end{subfigure}
         \hspace*{0.1 cm}
        \begin{subfigure}[b]{0.4\textwidth}
        	\centering
        	\begin{tikzpicture}[scale = 0.6]

        		\begin{axis}[legend pos =north west,ymin=0.0,ymax=600,xmin=0,xmax=2.0,xlabel= $ \rho_{B} (fm^{-3}) $,ylabel=$  P ( MeV/ fm^{3} ) $ ,axis on top,legend style={legend cell align=right,legend plot pos=right}]
        			
        			\addplot+[mark=none,
        			domain=0.1:0.5,
        			samples=100,
        			pattern=grid,
        			draw=blue,
        			area legend,
        			pattern color=blue]table{0.1lkw1.txt} \closedcycle;
        			\addlegendentry{Pure hadronic phase}
        			
        			\addplot+[mark=none,
        			domain=0.1:0.5,
        			samples=100,
        			pattern=dots,
        			draw=black,
        			area legend,
        			pattern color=black]table{0.1lkw2.txt} \closedcycle;
        			\addlegendentry{Mixed phase}
        			
        			\addplot+[mark=none,
        			domain=0.1:0.5,
        			samples=100,
        			pattern=crosshatch,
        			draw=orange,
        			area legend,
        			pattern color=orange]table{0.1lkw3.txt} \closedcycle;
        			\addlegendentry{Pure quark phase}
        			
        			\addplot[very thick,smooth,black,domain=0:2,samples=100] table{0.1lkw.txt};
        			\addlegendentry{Pressure}
        			
        		\end{axis}
        	\end{tikzpicture}
        	\caption{$\zeta=-0.1 $ combined with LKW}
        	\label{figef}
        \end{subfigure}
		\caption{Pressure vs baryon number density for hybrid matter within the Gibbs phase transition. Left (\subref{figea}, \subref{figec}, \subref{figee}): nuclear matter (NM) combined with the NJL model (RKH, HK, LKW), respectively. Right (\subref{figeb}, \subref{figed}, \subref{figef}): hadronic matter ($ \zeta = -0.1, -0.5, -0.1 $) combined with the NJL model (RKH, HK, LKW), respectively.}
		\label{fige}
	\end{figure}
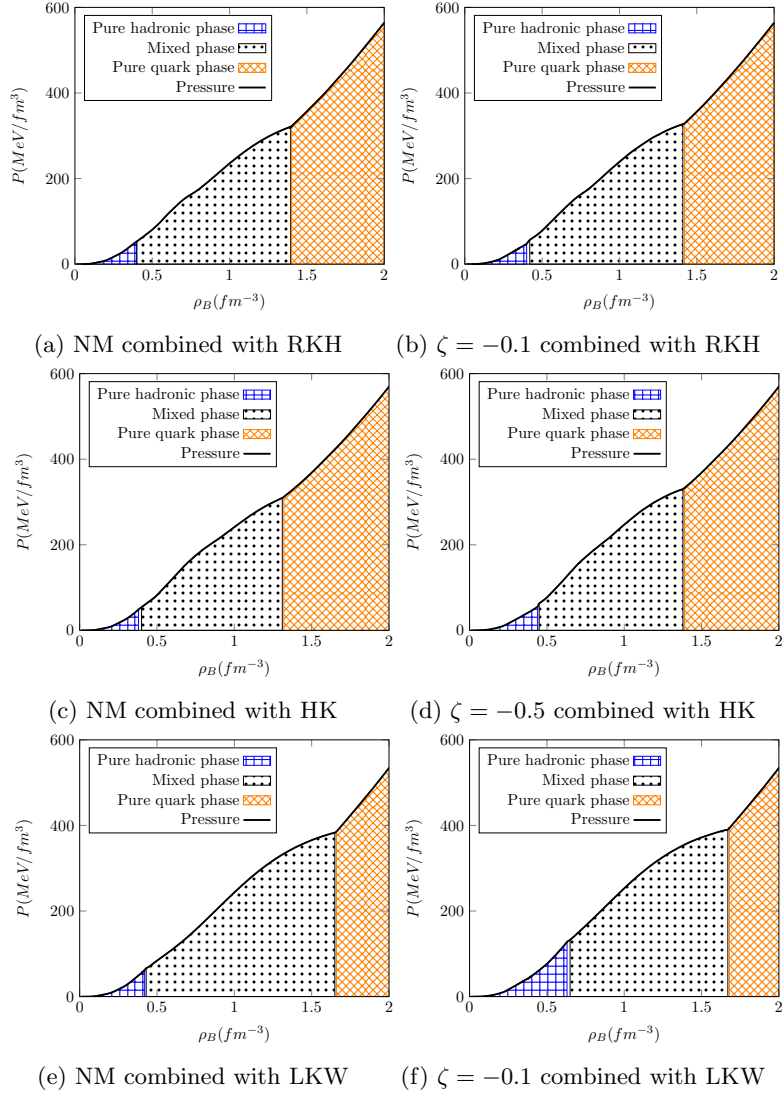

	In the final part of this section, the phase transition under Gibbs criteria is investigated for the NJL model. Also, the hadronic phase is the same as before. The equations of states, which include the hadronic phase, the mixed phase, and the quark phase, are represented in Fig. \ref{fige}.
	As shown in Figure \ref{fige}, the density at which the mixed phase region starts is almost the same for different parameter sets, $ \rho \approx 0.4 fm^{-3} $, except for \subref{figef}. However, the final density of the mixed phase is distinct for different parameter sets. Also, the maximum length of the mixed-phase regions is related to LKW parameter set (\subref{figee}, \subref{figef}), $\Delta\rho_{mixed} \approx 1.1 fm^{-3}$. Furthermore, by decreasing the value of $ \zeta $, the mixed phase area moves toward higher densities. Also, the results presented in the following sections show that the length of the mixed phase region does not have a direct effect on the maximum mass of the hybrid star. In fact, what is important is the $ \chi $ variable, which shows the ratio of the quark phase to the hadronic phase in the mixed phase region.
	
	The results of other studies that the hadronic phase described by RMF models are almost similar to our work. For example, when TM1 and IUFSU are used to describe the hadronic phase, the starting and ending points of the mixed phase are approximately equal to 0.5 and 1.5 $ fm^{-3} $, respectively \cite{z3}. On the other hand, when the hadronic phase is described by the BHF model, completely different results are obtained compared to our work, so that the beginning and end of the mixed phase are almost equal to 0.2 and 0.7 $ fm^{-3} $, respectively \cite{z4}.
	\subsection{Mass-radius relation of hybrid stars}
	In this section, the results related to the mass-radius calculations of the hybrid star are presented. The TOV equations are used to obtain the mass-radius relation of the star \cite{e07}. The TOV equations are:
	\begin{align}
		\frac{dP(r)}{dr}&=-\frac{GM(r)\epsilon(r)}{c^{2}r^{2}}(1+\frac{P(r)}{\epsilon(r)})(1+\frac{4\pi r^{3}P(r)}{M(r)c^{2}})(1-\frac{2GM(r)}{rc^{2}})^{-1},\\
		\frac{dM(r)}{dr}&=\frac{4\pi\epsilon(r)r^{2}}{c^{2}}.
	\end{align}
    where $M(r)$, $\epsilon(r)$ and $P(r)$ represent the mass, energy density and pressure of the hybrid star, orderly \cite{r3}. In the following, the results obtained from the mass-radius relation of the hybrid star for different quark models and distinct phase transitions are given.
	\subsubsection*{The MIT bag model (Maxwell criteria)}
\begin{figure}[h!]
		\centering
		\begin{subfigure}[b]{0.4\textwidth}
			\centering
			\begin{tikzpicture}[scale = 0.6]
				\begin{axis}
					[grid,legend pos =north east,ymin=0.0,ymax=2.2,xmin=9,xmax=25, ylabel=$ M/M_{sun} $,xlabel=$ R (km) $]
					\addplot[very thick,smooth,white,dashed] table{12.txt};
					\addplot[thick,smooth,blue,dashed] table{mrnpb80.txt};
					\addplot[thick,smooth,cyan,dotted] table{mrnpb100.txt};
					
					\addplot[thick,smooth,green,dashdotted] table{mrnpb130.txt};
					\addplot[thick,smooth,orange,loosely dotted] table{mrnpb150.txt};
					\addplot[thick,smooth,red,loosely dashed] table{mrnpb200.txt};

					\legend{NM+,B=80,B=100,B=130,B=150,B=200}
				\end{axis}
			\end{tikzpicture}
		\caption{ }
		\label{fig6a}
		\end{subfigure}
		\hspace*{0.1 cm}
		\begin{subfigure}[b]{0.4\textwidth}
			\centering
			\begin{tikzpicture}[scale = 0.6]
				\begin{axis}
					[grid,legend pos =north east,ymin=0.0,ymax=2.2,xmin=9,xmax=25, ylabel=$ M/M_{sun} $,xlabel=$ R (km) $]
					\addplot[thick,smooth,white,dashed] table{12.txt};
					\addplot[thick,smooth,blue,dashed] table{0.1mrb80.txt};
					\addplot[thick,smooth,cyan,dotted] table{0.1mrb100.txt};
					
					\addplot[thick,smooth,green,dashdotted] table{0.1mrb130.txt};
					\addplot[thick,smooth,orange,loosely dotted] table{0.1mrb150.txt};
					\addplot[thick,smooth,red,loosely dashed] table{0.1mrb200.txt};

					\legend{$ \zeta=-0.1 $ +,B=80,B=100,B=130,B=150,B=200}
				\end{axis}
			\end{tikzpicture}
		\caption{ }
		\label{fig6b}
		\end{subfigure}
	\hspace*{0.1 cm}
	\begin{subfigure}[b]{0.4\textwidth}
		\centering
		\begin{tikzpicture}[scale = 0.6]
			\begin{axis}
				[grid,legend pos =north east,ymin=0.0,ymax=2.2,xmin=9,xmax=25, ylabel=$ M/M_{sun} $,xlabel=$ R (km) $]
				\addplot[thick,smooth,white,dashed] table{12.txt};
				\addplot[thick,smooth,blue,dashed] table{0.5mrb80.txt};
				\addplot[thick,smooth,cyan,dotted] table{0.5mrb100.txt};
				
				\addplot[thick,smooth,green,dashdotted] table{0.5mrb130.txt};
				\addplot[thick,smooth,orange,loosely dotted] table{0.5mrb150.txt};
				\addplot[thick,smooth,red,loosely dashed] table{0.5mrb200.txt};

				\legend{$ \zeta=-0.5 $+,B=80,B=100,B=130,B=150,B=200}
			\end{axis}
		\end{tikzpicture}
	\caption{ }
	\label{fig6c}
	\end{subfigure}
    \caption{The mass-radius relation of hybrid star for the MIT bag model with different values of B combined with nuclear matter (NM) (\subref{fig6a}) and hadron matter with $ \zeta=-0.1 $ (\subref{fig6b}) and $ \zeta=-0.5 $ \subref{fig6c} within the Maxwell phase transition.}
    \label{fig6}
	\end{figure}
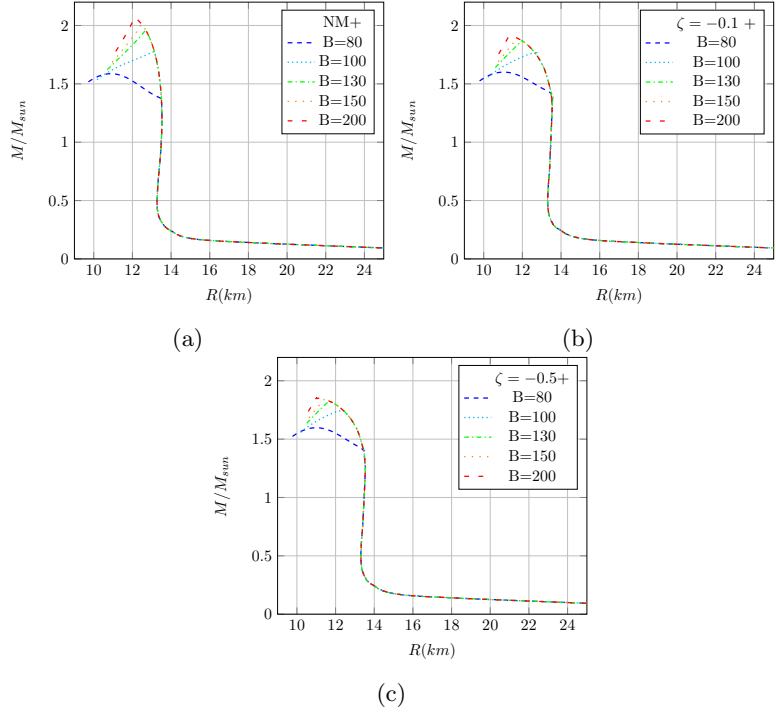

	First, we present the results related to the use of the MIT bag model within Maxwell's phase transition. These results are shown in Fig. \ref{fig6}.
	As shown in Fig. \ref{fig6a}, our calculations predict masses above \textbf{two} (2.07 $ M_{sun} $) for hybrid stars. But the maximum mass occurs at a point where the quark phase in the center of the hybrid star is not significant. In other words, a very thin layer of quark matter is predicted in the center of the star. Another point is that the maximum mass of the star also decreases with the decrease of the B value. Also, regardless of the type of equation of state used for the hadronic phase, when B is about 80 $ MeV/fm^{3} $, the maximum mass is about 1.6 $ (M_{sun}) $, and also there is  a considerable amount of quark matter in the star in this case (blue graphs in Fig. \ref{fig6})
	
	\subsubsection*{The MIT bag model (Gibbs's criteria)}
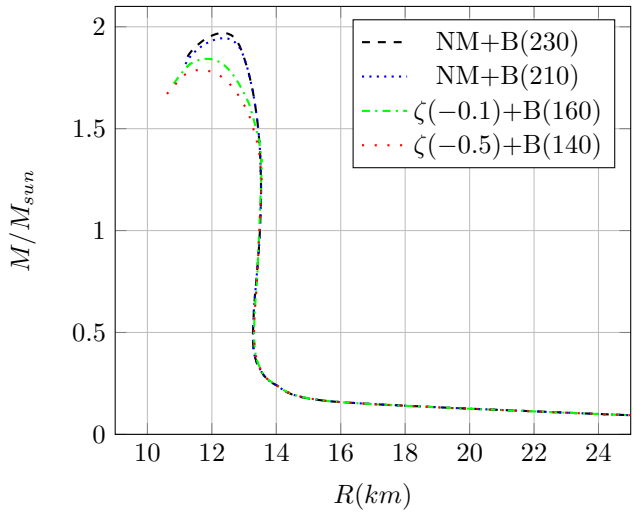
\begin{figure}[h!]
			\centering
			\begin{tikzpicture}
				\begin{axis}
					[grid,legend pos =north east,ymin=0.0,ymax=2.1,xmin=9,xmax=25, ylabel=$ M/M_{sun} $,xlabel=$ R (km) $]
					\addplot[thick,smooth,black,dashed] table{mr-np-b230.txt};
					\addplot[thick,smooth,blue,dotted] table{mr-np-b210.txt};
					
					\addplot[thick,smooth,green,dashdotted] table{mr-z0.1-b160.txt};
					\addplot[thick,smooth,red,loosely dotted] table{mr-z0.5-b140.txt};

					\legend{NM+B(230),NM+B(210),$ \zeta (-0.1)$+B(160),$ \zeta ( -0.5)$+B(140)}
				\end{axis}
			\end{tikzpicture}
		\caption{The mass-radius relation of hybrid star. Black: nuclear matter (NM) combined with the MIT bag model with B =230 $ (MeV /fm^{3}) $, Blue: nuclear matter (NM) combined with the MIT bag model with B =210 $ (MeV /fm^{3}) $, Green: hadronic matter with $ \zeta =-0.1 $ combined with the MIT bag model with B =160 $ (MeV /fm^{3}) $, Red: hadronic matter with $ \zeta =-0.5 $ combined with the MIT bag model with B =140 $ (MeV /fm^{3}) $, within the Gibbs phase transition.}
		\label{fig7}
	\end{figure}

	The results related to the phase transition under Gibbs conditions are shown in Fig. \ref{fig7}. Under the Gibbs phase transition, a maximum mass close to two $ (1.97 M_{sun}) $ is predicted for the hybrid star (The black graph in Fig. \ref{fig7}). Also, similar to before, the maximum mass of the star decreases by B or $ \zeta $. In these cases, each maximum mass occurs at a certain $ \chi $, which indicates the presence ratio of the quark phase in the hybrid star. As shown in Table \ref{tab3}, the lower the value of $ \chi $, the more maximum mass is predicted for the hybrid star.
    \begin{table}[h!]
    	\begin{center}
    		\begin{tabular}[h!]{|p{5cm}||p{3cm}||p{2cm}|}
    			\hline
    			\textbf{the EOS} & \textbf{Maximum mass $ M/M_{sun} $} & \textbf{$ \chi $}\\
    			\hline
    			\hline
    			NM combined with the MIT (B=230) & 1.969 & 0.1488\\
    			\hline
    			NM combined with the MIT (B=210) & 1.9435 & 0.1743\\
    			\hline
    			Hadronic matter ($ \zeta =-0.1$) combined with the MIT (B=160) & 1.8427 & 0.2653\\
    			\hline
    			Hadronic matter ($ \zeta = -0.5$) combined with the MIT (B=140) & 1.7879 & 0.3207\\
    			
    			\hline
    		\end{tabular}
    	\end{center}
    	\caption{The Maximum mass of the hybrid star for different EOS within the Gibbs phase transition, and also the corresponding $ \chi $ where the maximum mass of the star occurs.}
    	\label{tab3}
    \end{table}
    \subsubsection*{The NJL model (Maxwell criteria)}
\begin{figure}[h!]
    	\centering
    	\begin{subfigure}[b]{0.4\textwidth}
    		\centering
    		\begin{tikzpicture}[scale = 0.6]
    			\begin{axis}
    				[grid,legend pos =north east,ymin=0.0,ymax=2.1,xmin=9,xmax=25, ylabel=$ M/M_{sun} $,xlabel=$ R (km) $]
    				\addplot[very thick,smooth,white,dashed] table{12.txt};
    				\addplot[thick,smooth,blue,dashed] table{MR-NP-RKH.txt};
    				\addplot[thick,smooth,green,dotted] table{MR-Z0.1-RKH.txt};
    				
    				\addplot[thick,smooth,red,dashdotted] table{MR-Z0.5-RKH.txt};

    				\legend{RKH+,NM,$ \zeta=-0.1 $,$ \zeta=-0.5 $}
    			\end{axis}
    		\end{tikzpicture}
    	\caption{ }
    	\label{fig8a}
    	\end{subfigure}
    \hspace*{0.1 cm}
    \begin{subfigure}[b]{0.4\textwidth}
    	\centering
    	\begin{tikzpicture}[scale = 0.6]
    		\begin{axis}
    			[grid,legend pos =north east,ymin=0.0,ymax=2.1,xmin=9,xmax=25, ylabel=$ M/M_{sun} $,xlabel=$ R (km) $]
    			\addplot[very thick,smooth,white,dashed] table{12.txt};
    			\addplot[thick,smooth,blue,dashed] table{MR-NP-LKW.txt};
    			\addplot[thick,smooth,green,dotted] table{MR-Z0.1-LKW.txt};
    			
    			\addplot[thick,smooth,red,dashdotted] table{MR-Z0.5-LKW.txt};

    			\legend{LKW+,NM,$ \zeta=-0.1 $,$ \zeta=-0.5 $}
    		\end{axis}
    	\end{tikzpicture}
    \caption{ }
    \label{fig8b}
    \end{subfigure}
    \hspace*{0.1 cm}
    \begin{subfigure}[b]{0.4\textwidth}
    	\centering
    	\begin{tikzpicture}[scale = 0.6]
    		\begin{axis}
    			[grid,legend pos =north east,ymin=0.0,ymax=2.1,xmin=9,xmax=25, ylabel=$ M/M_{sun} $,xlabel=$ R (km) $]
    			\addplot[very thick,smooth,white,dashed] table{12.txt};
    			\addplot[thick,smooth,blue,dashed] table{MR-NP-HK.txt};
    			\addplot[thick,smooth,green,dotted] table{MR-Z0.1-HK.txt};
    			
    			\addplot[thick,smooth,red,dashdotted] table{MR-Z0.5-HK.txt};

    			\legend{HK+,NM,$ \zeta=-0.1 $,$ \zeta=-0.5 $}
    		\end{axis}
    	\end{tikzpicture}
    \caption{ }
    \label{fig8c}
    \end{subfigure}
    \caption{The mass-radius relation of the hybrid star for the NJL model for different parameter sets (RKH \subref{fig8a}, LKW \subref{fig8b}, HK \subref{fig8c}) combined with nuclear matter (NM) and hadronic matter with $ \zeta=-0.1 $ and $ \zeta=-0.5 $ within the Maxwell phase transition.}
    \label{fig8}
    \end{figure}
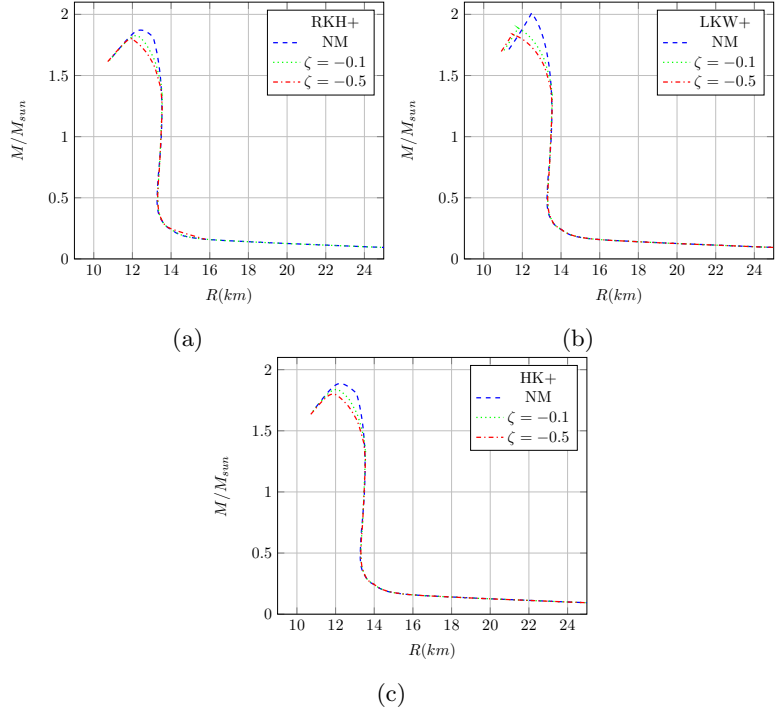

    In this section, we present the results related to the NJL model within Maxwell's phase transition. These results are shown in Fig. \ref{fig8}.
   As shown in Fig. \ref{fig8}, the maximum mass predicted for the hybrid star is related to LKW parameter set combined with nuclear matter, which is above \textbf{two} (2.01 $ M_{sun} $). But, in this case, a thin layer of quark matter is predicted in the center of the star. However, the highest maximum mass with a considerable amount of the quark phase is related to HK parameter set combined with nuclear matter (1.88 $ M_{sun} $) and RKH parameter set combined with nuclear matter (1.87 $ M_{sun} $), which predict a stable hybrid star. In all three parameter sets, the maximum mass of the hybrid star also decreases with the decrease of the $ \zeta $ value. For example, when $ \zeta $ decreases from -0.1 to -0.5, the maximum mass of the hybrid star also decreases from 1.82 to 1.80 ($ M_{sun} $).
   \subsubsection*{The NJL model (Gibbs's criteria)}
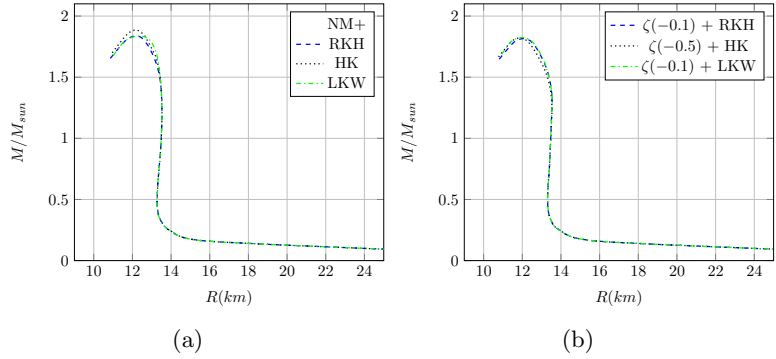
\begin{figure}[h!]
   	\centering
   	\begin{subfigure}[b]{0.4\textwidth}
   		\centering
   		\begin{tikzpicture}[scale = 0.6]
   			\begin{axis}
   				[grid,legend pos =north east,ymin=0.0,ymax=2.1,xmin=9,xmax=25, ylabel=$ M/M_{sun} $,xlabel=$ R (km) $]
   				\addplot[very thick,smooth,white,dashed] table{12.txt};
   				\addplot[thick,smooth,blue,dashed] table{mrnprkhe.txt};
   				\addplot[thick,smooth,black,dotted] table{mrnphke.txt};
   				
   				\addplot[thick,smooth,green,dashdotted] table{mrnplkwe.txt};

   				\legend{NM+,RKH,HK,LKW}
   			\end{axis}
   		\end{tikzpicture}
   		\caption{ }
   		\label{fig10a}
   	\end{subfigure}
    \hspace*{0.1 cm}
    \begin{subfigure}[b]{0.4\textwidth}
    	\centering
    	\begin{tikzpicture}[scale = 0.6]
    		\begin{axis}
    			[grid,legend pos =north east,ymin=0.0,ymax=2.1,xmin=9,xmax=25, ylabel=$ M/M_{sun} $,xlabel=$ R (km) $]
    			\addplot[thick,smooth,blue,dashed] table{0.1mrrkh.txt};
    			\addplot[thick,smooth,black,dotted] table{0.5mrhk.txt};
    			
    			\addplot[thick,smooth,green,dashdotted] table{0.1mrlkw.txt};

    			\legend{$ \zeta (-0.1) $ + RKH, $ \zeta (-0.5) $ + HK, $ \zeta (-0.1) $ + LKW}
    		\end{axis}
    	\end{tikzpicture}
    	\caption{ }
    	\label{fig10b}
    \end{subfigure}
    \caption{The mass-radius relation of the hybrid star for the NJL model for different parameter sets (RKH, LKW, HK) combined with nuclear matter (NM), \subref{fig10a}, and hadronic matter (HM), \subref{fig10b}, within the Gibbs phase transition.}
    \label{fig10}
   \end{figure}

   Finally, we present the results related to the NJL model within Gibbs's phase transition. These results are shown in Fig. \ref{fig10}.
   As shown in Fig. \ref{fig10}, similar to Maxwell's phase transition, the maximum mass predicted for the hybrid star with a considerable amount of the quark phase - with the corresponding $ \chi=0.619 $ where the maximum mass of the star occurs - is related to HK parameter set combined with nuclear matter, which is 1.88 $ M_{sun} $. Also, similar to before, the maximum mass of the star decreases by $ \zeta $.
   \section{Conclusion}\label{sec4}
   In the first part of this study, we investigated the possibility of the hadron-quark phase transition in the center of the hybrid star. To investigate this possibility, we used Maxwell and Gibbs's constructions. As mentioned before, the Sigma-omega-rho model - with correction on the hyperon coupling constants - was used to describe the equation of state of the hadronic part, and the MIT bag model and the NJL model were used to describe the quark phase. The phase transition occurs in almost all cases. However, the combination of the hadronic phase with the MIT bag model in the Gibbs phase transition and the combination of the hadronic phase with the NJL model in the Maxwell phase transition have a more logical prediction for the equation of states of the hybrid star. Because they predict a stable hybrid star with an acceptable amount of quark matter.
   
   Also, the combination of the hadronic phase with the MIT bag model in Maxwell's phase transition for small B has a good prediction of the equation of states of the hybrid star. For example, when B is equal to 80 $ (MeV /fm^{3}) $, a suitable phase transition occurs with a considerable amount of quark matter, and for larger B, only a thin layer of quark matter is predicted in the center of the hybrid star.
   
   The second topic that was studied in this article was obtaining the maximum mass of the hybrid star for different equation of states. In these calculations, we were able to predict the mass above 2 ($ M_{sun} $) for the star. However, in these cases, a noticeable quark phase in the center of the star is not predicted. The best predictions for the maximum mass of a stable hybrid star with a significant amount of quark phase are shown in Table \ref{tab4}.
    \begin{table}[h!]
    	\begin{center}
    	\begin{tabular}{|c||c|}
    		\hline
    		 \textbf{The EOS} & \textbf{The mass of the stable hybrid star}  ($ M_{sun} $) \\
    		 \hline
    		 \hline
    		 The MIT (Maxwell) & $ \approx 1.6 $ \\ 
    		 \hline
    		 The MIT (Gibbs) & $ \approx 1.9 $ \\ 
    		 \hline 
    		 The NJL (Maxwell) & $ \approx 1.9 $ \\ 
    		 \hline
    		 The NJL (Gibbs) & $ \approx 1.9 $ \\ 
    		 \hline
    	\end{tabular}
    \end{center}
    \caption{The maximum mass of the stable hybrid star with a considerable amount of quark matter.}
    \label{tab4}
    \end{table}

    As shown in Table \ref{tab4}, the NJL model predicts more maximum mass for the hybrid star than the MIT bag model. Also, when it comes to mass-radius relation of the hybrid star, the Gibbs phase transition has better results than the Maxwell phase transition. However, our results show that the Gibbs phase transition can not predict a pure quark phase at the center of the hybrid star. In other words, the maximum mass occurs in the mixed phase region, $ \rho_{max} $ is about $ 0.8 - 0.9 fm^{-3} $. In addition, as we expected, the $ \zeta $ parameter has a direct relationship with the maximum mass of the hybrid star, and the maximum mass increases with the increase of $ \zeta $.
    
    But the models we used to describe the quark phase also have some drawbacks. For example, the bag constant B and $ \Omega_{0} $ are constant for all densities in the MIT bag model and the NJL model, respectively. Therefore, if more accurate models are used to describe the quark phase, or if the drawbacks of the MIT bag model and the NJL model are removed \cite{e08,e09}, it is likely that you will get a more compatible equation of states for the hybrid star, which predicts more maximum mass for the star - probably near to the Tolman-Oppenheimer-Volkoff mass upper limit for neutron stars \cite{e10}.

\bibliographystyle{unsrt}
\bibliography{ref}
\end{document}